\documentclass{aa}
\usepackage{txfonts}
\usepackage{graphicx}
\usepackage{natbib}
\begin{document}

\title{Up to four planets around the M dwarf GJ 163}
\subtitle{Sensitivity of Bayesian planet detection criteria to prior choice}
\author{Mikko Tuomi\thanks{The corresponding author, \email{mikko.tuomi@utu.fi; m.tuomi@herts.ac.uk}}\inst{1,2} \and Guillem Anglada-Escud\'e\inst{3}}


\institute{
University of Hertfordshire, Centre for Astrophysics Research, Science and Technology Research Institute, College Lane, AL10 9AB, Hatfield, UK
\and University of Turku, Tuorla Observatory, Department of Physics and Astronomy, V\"ais\"al\"antie 20, FI-21500, Piikki\"o, Finland
\and Universit\"{a}t G\"{o}ttingen, Institut f\"ur Astrophysik, Friedrich-Hund-Platz 1, 37077 G\"{o}ttingen, Germany
}

\date{Received January 2013 / Accepted June 2013}

\abstract{Exoplanet Doppler surveys are currently the most efficient means to detect low-mass companions to nearby stars. Among these stars, the light M dwarfs provide the highest sensitivity to detect low-mass exoplanet candidates. Evidence is accumulating that a substantial fraction of these low-mass planets are found in high-multiplicity planetary systems. GJ 163 is a nearby inactive M dwarf with abundant public observations obtained using the HARPS spectrograph.}
{We obtain and analyse radial velocities from the HARPS public spectra of GJ 163 and investigate the presence of a planetary companions orbiting it. The number of planet candidates detected might depend on some prior assumptions. Since the impact of prior choice has not been investigated throughly previously, we study the effects of different prior densities on the detectability of planet candidates around GJ 163.}
{We use Bayesian tools, i.e. posterior samplings and model comparisons, when analysing the GJ 163 velocities. We consider models accounting for the possible correlations of subsequent measurements. We also search for activity-related counterparts of the signals we observe and test the dynamical stability of the planetary systems corresponding to our solutions using direct numerical integrations of the orbits.}
{We find that there are at least three planet candidates orbiting GJ 163. The existence of a fourth planet is supported by the data but the evidence in favor of the corresponding model is not yet conclusive. The second innermost planet candidate in the system with an orbital period of 25.6 days and a minimum mass of 8.7 M$_{\oplus}$ is inside the liquid-water habitable zone of the star.}
{The architecture of GJ 163 system resembles a scaled-down Solar System in the sense that there are two low-mass planets on orbital periods of 8.7 and 25.6 days in the inner system, a possible slightly more massive companion on an intermediate orbit, and an outer sub-Saturnian companion at roughly 1 AU. The discovery of (yet) another planetary system with several low-mass companions around a nearby M-dwarf indicates that the high-multiplicity planetary systems found by the NASA Kepler mission around G and K dwarfs is also present (possibly even reinforced) around low-mass stars.}

\keywords{Methods: Statistical, Numerical -- Techniques: Radial velocities -- Stars: Individual: GJ 163}



\maketitle


\section{Introduction}

Observing the periodic Doppler signatures of planetary companions in the stellar spectra is currently the most efficient method of finding exoplanets around nearby stars. This radial velocity (RV) method can be readily applied to M dwarfs that are the most populous stars in the Solar neighbourhood and are known to be hosts to several planetary systems, e.g. GJ 581 \citep{bonfils2005,udry2007,mayor2009}, GJ 667C \citep{anglada2012,delfosse2012,bonfils2013}, GJ 676A \citep{forveille2011,anglada2012b}, and GJ 876 \citep{delfosse1998,marcy1998,marcy2001,rivera2005,rivera2010}. Because of the several on-going high-precision RV surveys, this population of exoplanet systems with multiple planets can readily be expected to increase as the observational baselines extend and as the amount of high-precision data increases and enables the detections of signals with lower and lower amplitudes.

In this article, we apply Bayesian multi-planet detection techniques \citep[e.g.][]{tuomi2012c,tuomi2012d} to analyse new RV measurents obtained with the HARPS-TERRA software \citep{anglada2012c} from public HARPS spectra of the nearby M dwarf GJ 163\footnote{As requested by the anonymous referee, we note that in a talk given by T. Forveille at the IAU XXVIII General Assembly in Beijing, August 2012, it was announced that GJ 163 is orbited by at least two planets (H. R. A. Jones, priv. comm.). However, according to our knowledge, a correponding study has not been made public and thus we could not verify neither this claim nor the nature of the proposed system.}. Since a number of subjective choices are required by any Bayesian statistical data analyses, we discuss how this subjectivity -- i.e. prior probability densities or prior models -- affects our detection sensitivity and reliability. In particular, we analyse the same dataset by assuming different prior choices and noise models in order to obtain as objective results as possible.

Since the nature of RV noise of main sequence stars is not very well known and some stars show clear autocorrelation in their velocity noise \citep{tuomi2012d,tuomi2012e,baluev2012}, we analyse the GJ 163 velocities using different noise models, i.e. Gaussian white noise and a simple red noise model. We perform the analyses using posterior samplings \citep[following ][]{tuomi2012,tuomi2012b,tuomi2012c} and calculate estimates for the Bayesian evidences using the truncated posterior mixture estimate \citep{tuomi2012b}. As a result, we report our best estimates for the number and properties of planetary signals in the RV data of GJ 163 using the different prior and noise model choices. As a final validation procedure, we also investigate whether these signals correspond to any clear periodic features in the activity data of the star.

Prior probability densities are an integral part of any statistical analyses based on the Bayes' rule of conditional probabilities. However, most, if not all, frequentist methods can in fact be derived from Bayesian ones by making certain simplifying assumptions on the shapes and natures of the prior probability densities and likelihood functions of the model parameters and measurements. For instance, all the maximum likelihood methods (e.g., the popular multi-variate least-squares minimisation methods) can be derived from the Bayesian ones by adopting uniform prior probability densities for the model parameters over some suitable range. Similarly, all the confidence level tests based on $\chi^{2}$ statistics can be derived from the Bayes' rule and Bayesian evidence ratio tests by assuming Gaussian likelihood functions and uniform prior densities. Because of its generality, only Bayesian framework enables to study the effects these subjective choices might have on the obtained results. All statistical analysis methods contain prior information (uniform prior is a prior as well) but only Bayesian ones enable taking them into account in a logically consistent manner.

When searching for planetary signatures in noisy data using the Doppler method, the choice of prior probability densities has not been discussed very extensively in the literature \citep[some examples can be found in][]{baluev2012,tuomi2012b}. \citet{ford2007} defined a set of rather uninformative prior densities for analyses of radial velocity data using the typical statistical models consisting of Keplerian signals and white noise. These priors were modified slightly in \citet{tuomi2012} and \citet{tuomi2012c} but the effects different prior choices have on the obtained results have not been studied. We study these effects in this article by comparing the effects of prior models on the obtained results when analysing the RV data of GJ 163. We discuss and motivate our prior choices in Section \ref{sec:priors} and present the full analysis results of GJ 163 HARPS-TERRA velocities in Section \ref{sec:gj163_analysis}.

\section{Prior choice}\label{sec:priors}

The prior probability densities of model parameters represent the other component of Bayesian statistical tools in addition to the likelihood functions. The need to compare different likelihood functions, or likelihood models, is understood to be an important feature of statistical analyses of noisy data when the attempt is to find out the processes producing the observations, but the prior probability densities play a significant role as well and different prior models warrant comparisons in order to find the most trustworthy descriptions for the data.

While it is typically the case that the likelihood ``overwhelms'' priors in the sense that the likelihood function sets much stronger constraints to the posterior density than the prior, this is not necessarily such a good idea when there are not many more measurements than free parameters and when the noisy data does not constrain the parameters much. In such cases, it might be necessary to use stronger prior constraints, i.e. informative priors, to obtain any sensible solutions at all. Uninformative prior choices are generally preferred because such prior beliefs have as little effect as possible on the obtained results. In general, this is important because the relevance of a result is usually strongly tied to how constraining the prior choices are in practice, i.e. what are the assumptions (implicit or explicit) enabling one to reach a significant conclusion. In the next subsection, we will show that even uninformative priors can have, in principle, considerable effects on the obtained results and the inferred conclusions.

\subsection{Uninformative priors}

When performing statistical analyses of noisy data, the goal is typically to obtain as objective results as possible. In practice, the prior beliefs of a scientist performing data analyses should not be allowed to affect the obtained results -- especially if they differ radically from the prior beliefs of fellow scientists. Priors that all scientists can consider rather objective are usually called \textit{uninformative priors} because they aim at describing a maximum amount of ignorance on the system that the data is assumed to describe.

Uniform probability densities are a common example of such uninformative priors. Generally, for parameter $\theta$, they can be written as $\pi(\theta) = c$ for all $\theta \in \Omega$, where $\Omega$ is the parameter space that is usually assumed to be a bounded subset of the real line to make the density a proper probability density and $c$ some positive constant. However, because analysis results cannot depend on the chosen unitary system and should remain independent after any linear transformation of the parameters, it is always possible to choose $c=1$.

Uniform priors cannot be considered uninformative for all the parameters, especially so, when searching for periodic signals in noisy data corresponding to signatures of extra-solar planets. For instance, it is unrealistical to assume \emph{a priori} that an RV data set has an equal probability of containing a periodic signal of planetary origin between 0-100 days as between 100-200 days because the former contains much more orbits that are stable than the latter. Therefore, period ($P$) is a scale-parameter and an invariant prior for such a parameter is the Jeffreys' prior that can be written as $\pi(P) \propto P^{-1}$. This prior corresponds to a uniform prior probability density in the log-period space, which is the rationale behind the common choice of $\log P$ as a parameter in the statistical model instead of $P$.

\subsection{Transformation of parameters}

Under a nonlinear transformation from parameter vector $\theta$ to vector $\theta'$, the corresponding changes in prior densities have to be accounted for in order to obtain consistent results using both parameterisations. Generally, a transformation $\theta \rightarrow \theta'$ results in a change in the prior probability density defined as
\begin{equation}\label{eq:prior_transformation}
   \pi(\theta') = \pi(\theta) \bigg| \frac{d \theta}{d \theta'} \bigg| ,
\end{equation}
given that the transformation $\theta' = f(\theta)$ has an inverse transformation and $| \frac{d \theta}{d \theta'} |$ is the Jacobian of this transformation. It is easy to see that a linear transformation of the form $\theta' = a\theta + b$ for $\pi(\theta) = c$ yields $\pi(\theta') = c'$. This means that under a linear transformation, i.e. a change of unit system, constant priors remain unchanged. However, this is not the case in general, which means that for any non-linear transformation between $\theta$ and $\theta'$, the results might be rather different if uniform priors were used in both parameterisations.

\subsection{Candidate priors}

In this subsection, we define the candidate prior probability densities we use to analyse the velocities of GJ 163.

As a set of reference priors, we use those in presented in \citet{tuomi2012} because this choice is a rather uninformative one and does not constrain the model parameters strongly \citep[e.g.][]{anglada2012b,tuomi2012,tuomi2012c}. The functional forms of the priors and the limits of the respective parameter spaces of the model parameters are shown in Table \ref{tab:ref_priors}. Because all the parameters are real numbers, we denote the parameter spaces as intervals of the real line. The parameters of the Keplerian signals in Table \ref{tab:ref_priors} are the radial velocity amplitude ($K$), longitude of pericentre ($\omega$), orbital eccentricity ($e$), mean anomaly ($M_{0}$), and the logarithm of the orbital period ($l = \log P$). In addition to these Keplerian parameters, we also include a constant reference velocity ($\gamma$), a radial velocity jitter ($\sigma_{J}$) (amount of excess white noise in the data on top of the estimated instrument uncertainties), and a correlation coefficient between the noise of subsequent epochs ($\phi$) as defined in e.g. \citet{tuomi2012c} and \citet{tuomi2012d}.

\begin{table}
\center
\caption{Reference prior probability densities of and ranges of the model parameters\label{tab:ref_priors}}
\begin{tabular}{lcc}
\hline \hline
Parameter & $\pi(\theta)$ & Interval \\
\hline
$K$ & Uniform & $[0, K_{max}]$ \\
$\omega$ & Uniform & [0, $2 \pi$] \\
$e$ & $\propto \mathcal{N}(0, \sigma_{e}^{2})$ & [0,1] \\
$M_{0}$ & Uniform & [0, $2 \pi$] \\
$l$ & Uniform & $[\log P_{0}, \log P_{max}]$ \\
$\sigma_{J}$ & Uniform & $[0, K_{max}]$ \\
$\gamma$ & Uniform & $[-K_{max}, K_{max}]$ \\
$\phi$ & Uniform & [-1, 1] \\
\hline \hline
\end{tabular}
\end{table}

We choose the hyperparameters of each prior distribution in Table \ref{tab:ref_priors} as follows.

\subsubsection{Semi-amplitude}

The maximum semi-amplitude is set to $K_{max} = $20 ms$^{-1}$. This choice is motivated by the fact that the standard deviation of the RVs is 6.8 ms$^{-1}$ and, therefore, we do not expect to find signals that have amplitudes in excess of this maximum amplitude.

\subsubsection{Eccentricity}

One of the most disputed and controversial priors choices in the statistical analysis of the Keplerian problem is the prior choice for the eccentricity. In this study, we consider three cases by using three choices of the hyperparameter $\sigma_{e}$ (Table \ref{tab:ref_priors}). As in previous works \citep[e.g.][]{tuomi2012}, our reference choice is $\sigma_{e} = 0.3$ but we also obtain results with values of $0.2$ and $0.1$.

\begin{figure}
\center
\includegraphics[width=0.43\textwidth]{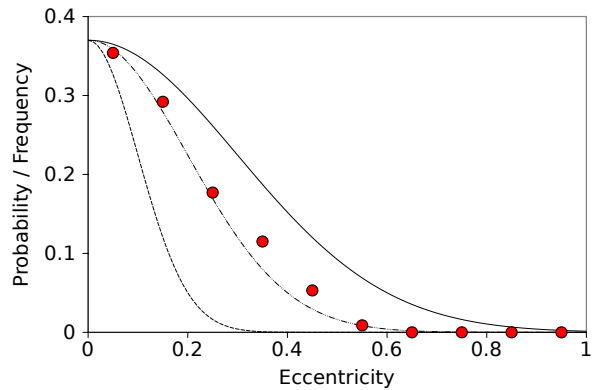}
\caption{The eccentricity distribution (red circles) of low-mass ($m_{p} < 0.1$M$_{\rm Jup}$) exoplanet candidates and the relative eccentricity priors with $\sigma_{e} = 0.3$ (solid line), $\sigma_{e} = 0.2$ (dash-dotted line), and $\sigma_{e}= 0.1$ (dashed line). The priors have been scaled to have maxima of 0.37 to enable visual comparison with the distribution.}\label{fig:ecc_freq}
\end{figure}

The low value of $\sigma_{e} = 0.1$ can be justified in the following way. The hyperparameter $\sigma_{e}$ controls how we expect the probability of the eccentricity to decrease as it approaches unity. Our reference prior might not be of very practical use when the RV data-set is relatively small, it has uncertainties comparable to in magnitude to the Keplerian signals it might contain, and its sampling cadence is very uneven (e.g. long gaps). In such cases, too uninformative assumptions, i.e. too high values of $\sigma_{e}$, would allow posterior probability densities to have roughly equally high values over large subsets of the parameter space making the interpretation of the obtained results very difficult, if possible at all. In the Keplerian problems, it is a common trick to restrict the parameter space of low amplitude candidates to strictly circular orbits \citep[e.g.][]{lovis2011,pepe2011,dumusque2012}, which decreases the number of free parameters in the model by $2\times N_{\rm planets}$. This assumption also avoids the strong non-linear regime of highly eccentric orbits where the implicit assumptions of point estimate methods (e.g., multivariate Gaussian posteriors) do not hold anymore. This choice corresponds to a delta-function prior density of the form $\pi(e) = \delta(e)$ that gives all the prebability density to $e = 0$. We considered this choice to be too limiting and selected a slightly less informative one, i.e. $\mathcal{N}(0, \sigma_{e}^{2})$ with $\sigma_{e} = 0.1$. This hyperparameter results in a prior density that is much narrower than the observed eccentricity distribution of low-mass planets (Fig. \ref{fig:ecc_freq}) but not as limiting as an eccentricity fixed to zero.

To investigate how the current statistical properties of the observed exoplanet candidates support our choice of the eccentricity prior, we obtained data from The Extrasolar Planets Encyclopaedia\footnote{The web site \texttt{exoplanet.eu} maintained by J. Schneider.} and selected a sub-sample of all listed exoplanets with minimum masses lower than 0.1 M$_{\rm Jup}$. As obtained in Dec 6th, 2012, this sample contained 113 planet candidates for which eccentricity has been estimated. For 22 of them the reported estimate was exactly zero, which suggests that their eccentricities have been fixed to zero in the statistical analyses, or that only an upper limit was provided at the time of publication. Nevertheless, we assume that the eccentricities of these 22 candidates are indeed close to zero and put them in the $[0-0.1]$ bin in Fig. \ref{fig:ecc_freq}. The obtained eccentricity distribution is illustrated in Fig. \ref{fig:ecc_freq} where we also add a representation of our three test cases (eccentricity priors with $\sigma_{e} = $0.1, 0.2 and 0.3). This figure shows that -- for the sample of low-mass planets -- the eccentricity prior with $\sigma_{e} = 0.2$ appears to be a reasonably good representation of the observed population, and therefore, it also poses an interesting choice for a prior density to be investigated.

We note that the eccentricities in the literature are likely overestimated in general for the simple reason that eccentricity is a positive number and cannot be lower than zero. This means that if the eccentricity estimate of a low-eccentricity orbit is biased, it it necessarily more likely to be overestimated than underestimated because the former would be more likely than the latter \citep{zakamska2011}. Eccentricities are also overestimated systematically when the signal amplitude is not much larger than the measurement noise and when the number of observations w.r.t. the model parameters is low \citep{otoole2008,shen2008}. Therefore, our eccentricity prior might be a reasonable choice even for $\sigma_{e} = 0.1$.

\subsubsection{Jitter}

Because we might know that the target star is a quiescent and a rather inactive one (as is the case of GJ 163, see Section \ref{sec:activity}), we assign an informative prior density to the stellar jitter. A fixed jitter parameter is sometimes chosen based on reasonable, yet subjective, considerations such as maximum demonstrated precision of an instrument, or observed typical variability of other similar stars without planets. As for the eccentricity prior on the circular orbital case, it is a common practice to assume a rather restrictive and informative $\delta$-function prior density for the jitter, i.e. to fix the jitter parameter to some \emph{a priori} estimated value. To relax this condition and still be able to use our prior knowledge of the sample, we chose $\pi(\sigma_{J}) \propto \mathcal{N}(\mu_{\sigma}, \sigma_{\sigma}^{2})$ and selected the hyperparameters as $\mu_{\sigma} = 1$ ms$^{-1}$ and $\sigma_{\sigma} = $1 ms$^{-1}$. These estimates are based on a sample of 27 M dwarfs targeted by HARPS that was found to have on average 1.9 ms$^{-1}$ variation (after removing the signals of all known planet candidates) and only two of them had variations in excess of 4 ms$^{-1}$ \citep{dasilva2012}. Because effectively we add this parameter describing the stellar jitter in quadrature to the estimated instrument uncertainties (Table \ref{tab:rvs}) in our model, we consider that our prior choice represents the observed properties of M dwarfs reasonably well.

\subsubsection{Period}

The minimum and maximum periods in the parameter space are chosen as $P_{0} =$ 1 day and $P_{max} = 2T_{obs}$, where $1$ day is the typical separation between consecutive observations and $T_{obs}$ is the baseline of the observations\footnote{This kind of dependence of the hyperparameters on the data is sometimes referred to as ``data dependent prior''. It is not a prior in the traditional sense because it is not completely independent of the data but is rather commonly used in statistical literature.}.

Especially when dealing with periodic signals, the period is the most important parameter in terms of confidently discovering a signal in the first place. However, because the parameter we actually use in our posterior samplings is not the period as such but its logarithm, we must note that the actual prior in the period space is then obtained by using the Jacobian of the transformation (Eq. \ref{eq:prior_transformation}). After simple algebra it can be seen that our reference prior choice (uniform in $\log P$) corresponds to an implicit prior of the period parameter of the form $\pi(P) \propto P^{-1}$. This type of dependence might not seem to be very satisfactory one as a prior because it assigns higher probabilities for shorter periods than for longer ones. However, it corresponds to a scale-invariant prior, i.e. the Jeffreys' prior, that remains the same regardless of the unit system of the period parameter. A uniform prior in frequency corresponds to an implicit prior of $\pi(P) \propto P^{-2}$ and is essentially the one used when analysing time series in the frequency space using periodogram based methods \citep{baluev2012}. Therefore, we compare the three prior choices that seem natural to the problem: uniform in $\log P$, uniform in $P$, and uniform in $P^{-1}$.

As already mentioned, our sampling strategy uses the logarithm of the orbital period as a parameter in the posterior samplings because it is a scale-invariant parameter. Therefore, we must express the priors uniform in $P$ and in $P^{-1}$ in the log-orbital period space. Simple application of Eq. (\ref{eq:prior_transformation}) shows that for $\pi(P) = 1$ it follows that $\pi(l) \propto \exp(l)$. Similarly, for $\pi(P^{-1}) = 1$ it follows that $\pi(l) \propto \exp(-l)$. Clearly, the differences between these priors and a uniform prior in $l$ are considerable and certainly can be expected to have an effect on the obtained results. Yet, without using physical constrains, it cannot be said that any of them is more wrong than the others. While our reference prior of choice is the uniform one in $\log P$, we study the effect of the alternative ones on the interpretation of the results obtained from the velocity data in Section \ref{sec:gj163_analysis}.

\section{GJ 163}\label{sec:star}

GJ 163 is a nearby M3.5 V dwarf \citep{koen2010} with a Hipparcos parallax of 66.69$\pm$1.82 mas \citep{vanleeuwen2007}, which implies a distance of $\sim$ 14.9$\pm$ 0.4 pc. We use this distance with J, H and K photometry from 2MASS \citep{cutri2003} to derive a mass of 0.40$\pm$ 0.02 $M_{\odot}$ using the mass-luminosity relation given by \citet{delfosse2000}. Since the uncertainty in the distance is rather low, the estimate of the mass is mostly limited by the precision of the calibration ($\sim$ 5\% for M$_{\star}<0.5 $M$_{\odot}$). Using the photometric metallicity calibrations from \citet{johnson2009} and \citet{schlaufman2010} and the V and K magntidues of the star, we obtain a slightly super-solar metallicity of [Fe/H] $=$ 0.1 $\pm$ 0.1 (mean of the two calibrations), where the uncertainty is again intrinsic uncertainty in the empirical relations. This is the metallicity range where the evolutionary models produce a better agreement with observations. Using \citet{chabrier1997} and assuming an age between 2 and 10 Gyr, we obtain a luminosity of 0.0196 L$_{\odot}$ and a corresponding effective temperature of 3500 $\pm$ 100 K.

\begin{figure}[tb]
   \centering
   \includegraphics[width=0.45\textwidth,clip]{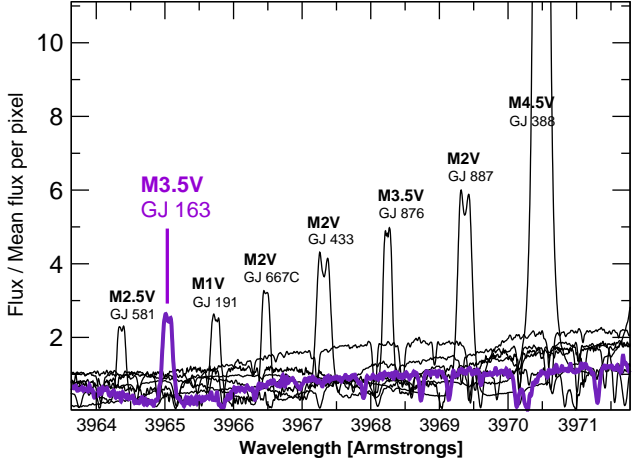}

\caption{CaII H line of GJ 163 (purple) as obtained from co-adding the 55 available HARPS spectra. The same lines of stars with similar spectral type are shown for comparison (offset in wavelength). The height of the line was scaled to the mean number of counts per pixel in this particular HARPS echelle order. Spectral types were taken from \citet{bonfils2013}. As discussed in the text, GJ 163 is among the least chromospherically active stars in its spectral range.} \label{fig:cakline} 
\end{figure}

The parallax and proper motion of GJ 163 imply a UVW velocity vector of (-70.1, -75.5, 0.51) kms$^{-1}$ which has rather high components for a typical thin disk star, meaning that the star is more likely a member of the thick disk \citep[see Figure 3 in][]{bensby2003}. Compared to other M dwarfs of similar spectral class, GJ 163 has a rather low S-index (0.61) -- between the very stable and planet prolific GJ 581 (0.46) and the slightly more ``jittery'' but also planet prolific GJ 876 (0.82). The relative heights of the CaII K lines of some typical M dwarfs are shown in Fig. \ref{fig:cakline}. This relatively low emission in CaII indicates that GJ 163 has low activity levels and suggests that it is rather old ($>2$ Gyrs).

\begin{table}
\caption{Basic parameters and derived properties of GJ 163}
\label{tab:star}
\centering                                      
\begin{tabular}{lcc}
\hline \hline
Parameter               & Value                 & Reference \\
\hline
R.A.                    &  04 09 15.663         & (a) \\
Dec.                    & -53 22 25.305         & (a) \\
$\mu^*_{\rm R.A.}$
$[$mas yr$^{-1}]$       & 1041.43 $\pm$  2.0    & (a) \\
$\mu_{\rm Dec}$
[mas yr$^{-1}$]         & 583.01  $\pm$  2.0    & (a) \\
Parallax [mas]          &   66.69 $\pm$  1.62   & (a) \\
V                   & 11.811  $\pm$  0.01   & (b) \\
J                   & 7.948   $\pm$  0.03   & (c) \\
H                   & 7.428   $\pm$  0.04   & (c) \\
K                   & 7.135   $\pm$  0.02   & (c) \\
Sp. type          & M3.5 V                 & (b) \\  
Mass [M$_\odot$]    & 0.4     $\pm$ 0.02    & (d) \\    
Fe/H                    & +0.1    $\pm$ 0.10    & (e) \\    
Mean S-index            & 0.61    $\pm$ 0.01    & (f) \\    
T$_eff$  [K]            & 3500    $\pm$ 100     & (g) \\
L$_{\star}$  [L$_\odot$]      & 0.0196  $\pm$ 0.001   & (g) \\
\hline \hline
\end{tabular}
\tablefoot{(a) Hipparcos catalog, \citet{vanleeuwen2007}; (b) \citet{koen2010}; (c) 2MASS catalog, \citet{cutri2003}; (d) Using \citet{delfosse2000}; (e) Using the mean of \citet{johnson2009} and \citet{schlaufman2010}; (f) this work; (g) this work using \citet{chabrier1997}.}
\end{table}

Given the precision in the distance and in the photometric colours, the estimation of the total luminosity we provide is dominated by systematic model uncertainties at the level of $\sim$ 5\% \citep{boyajian2012}. While empirical calibrations based on interferometric radius measurements of some M dwarfs exist, they still result in large scatter ($\sim$ 20--30\%) at effective temperatures below 4000 K \citep{boyajian2012}. Therefore, and besides all caveats associated with using models, we believe that \citet{chabrier1997} provides a more reliable estimate of the stellar luminosity and we assign a 5\% uncertainty to it.

\subsection{Observations}

\begin{figure}
\center
\includegraphics[angle=-90,width=0.4\textwidth]{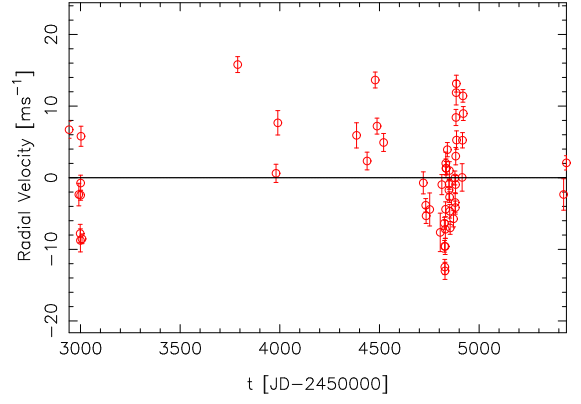}
\caption{HARPS-TERRA radial velocities of GJ 163 with the data median removed.}\label{fig:rvs}
\end{figure}

The radial velocities underlying the current work are based on spectra obtained with the HARPS spectrograph during different observing programs over the recent years\footnote{Programs: 072.C-0488/PI-M.Mayor,  082.C-0718/PI-X. Bonfils and 085.C-0019/PI-G. Lo Curto}. Visual inspection of the RV measurements shown in Fig. \ref{fig:rvs} shows that the star shows Doppler variability in excess of the typical uncertainties ($\sim$ 1.1 m s$^{-1}$) from the very first observations (obtained in 2003). A more intensive set of high-cadence observations was obtained in 2009 in the context of the HARPS-Totems program (PI. X. Bonfils) whose aim was to detect short period super-Earth- and Neptune-mass planet candidates with high probability of transit. According to the HARPS-ESO archive\footnote{http://archive.eso.org/wdb/wdb/eso/repro/form}, the star is still being monitored but observations made after August 2010 have not been made available.

The HARPS spectra are taken with typical exposure times of 900 seconds and the median signal-to-noise ratio (S/N) per pixel at 6100 \AA ~ is $\sim$ 35. While this might not seem very high, spectra of M-dwarfs contain considerable amounts of Doppler information in the redder end of HARPS. This information is optimally extracted by the HARPS-TERRA software which consist on matching the extracted spectrum as provided by the HARPS-DRS (data reduction software) to a very high S/N template made by coadding all the observations. The HARPS-DRS spectra are generated from a standardarized calibration set of observations that include dark, flat fielding and calibration lamp exposures (Th-Ar lamp) obtained at the beginning of each night. These sets of calibrations together with a carefully designed calibration plan have ensured the wavelength solution consistency down to a precision of 1 ms$^{-1}$ or better over the years \citep[e.g.][]{lovis2007}. Given that there were a total of 55 spectra available, the combined template spectrum has a S/N ratio of $\sim$ 250 at 6100 \AA ~ contributing negligibly to the error budget of each individual measurement. The high precision achievable with the HARPS-TERRA processing has been demonstrated in \citet{anglada2012c} and its comparisons to the HARPS-DRS derived velocities (obtained via cross correlation with a weighted binary mask) have generally shown that the HARPS-TERRA velocities have greater precision for M dwarfs \citep{anglada2012,anglada2012b,anglada2013}. In these studies, it has also been shown that some of the HARPS M-dwarf sample stars are as stable (even more stable, with RMS $<$ 1 m s$^{-1}$ over several years) than some other primary targets from the HARPS-GTO sample \citep[mid-K to late G-dwarfs, such as HD 88512 or Tau Ceti][]{pepe2011}. The Doppler time-series, as well as measurements of activity indices, are given in Table \ref{tab:rvs} \citep[see][for further details]{lovis2011,anglada2012c}.

\section{Signals in activity indices} \label{sec:activity}

In addition to random noise, stellar activity can also generate apparent Doppler periodicities that can be confused with true Keplerian signals. Before going into the analysis of the Doppler data, we first investigate whether there are periodic variations in the three representative activity indices of GJ 163 \citep[see e.g.][]{anglada2012b}. These indices are the S-index, the line bisector (BIS) and the full-with at half-maximum (FWHM) of the cross correlation function.

The S-index is proportional to flux coming from the chromosphere of the star. It is measured as the amount of emission at the Ca II H+K doublet (393.3664 nm for the K and 396.8470 nm for the H line) compared to a locally defined continuum. The recipes for computing the S-index from HARPS spectra are given in \citet{lovis2011} and \citet{anglada2012c} describe how HARPS-TERRA obtains them from the HARPS-DRS products. Such chromospheric emission is closely related to the intensity of the stellar magnetic field and larger values imply higher activity levels. The S-index can show periodic variations due to the stellar magnetic cycle (e.g. 11-years magnetic cycle of the Sun), episodic events of higher activity (e.g. flares) and due to the presence of active regions on the rotating surface of the star. The FWHM and the BIS monitor changes in the mean spectral line profiles and should correlate with the presence of spots. Temperature contrast spots and/or magnetic spots are known to show correlation with spurious Doppler signals \citep[e.g.][]{queloz2001,reiners2013}.

Our general strategy is as follows. If a strong periodicity is identified in any of the indices and such periodicity could be related to any of the Doppler signals (compatible period or related through first order aliases), we add a linear correlation term to the model of the Doppler data and perform new samplings of the parameter space. If the data were better described by the correlation term rather than a genuine Doppler signal, the overall model probability would increase and the planet signal in question should disappear (or its amplitude would be altered substantially).

\begin{figure}
\center
\includegraphics[width=0.45\textwidth,clip]{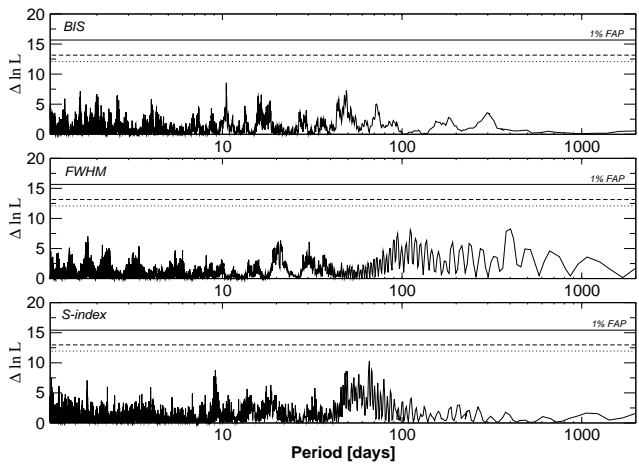}
\caption{Log-likelihood periodograms of HARPS activity indicators of GJ 163: BIS (top panel), FWHM (middle panel), and $S$-index (bottom panel).\label{fig:activity}}
\end{figure}

For GJ 163 in particular, none of the activity-indices shows any hints of periodic variability at all. This lack of activity is also supported by the fact that GJ 163 has a mean value of the S-index comparable to the ones measured on the planet prolific M dwarfs GJ 581 and GJ 667C. As a note, we neglected 4 FWHM and BIS measurements due to manifestly incorrect estimates produced by the HARPS-ESO data-reduction software (see Table \ref{tab:rvs}). All 55 measurements of the S-index were successfully extracted by the HARPS-TERRA software and used in the analysis. We show the periodograms of the three activity-indices in Fig. \ref{fig:activity}.

To search for signals in the indices, we used log-likelihood periodograms, which are computationally substantially less intensive and require much less ``supervision'' than Bayesian Markov chains. These periodograms represent the improvement of the likelihood of the model \citep[log-likelihood periodograms,][]{baluev2009} against the null hypothesis for each test period. Compared to more classic periodograms based solely on $\chi^2$ minimization \citep[e.g.][]{cumming2004}, our log-likelihood periodogram also adjusts the white noise component of the noise. As any generic periodogram approach, if the model including a periodic signal (sinusoid) substantially improves the merit statistic (likelihood function in this case), a peak over the 1\% false alarm probability threshold should emerge. For GJ 163 in particular, none of the indices shows any consistent periodicity at all. For each index, we also attempted to search for a second signal but did not find any significant improvements to the model. In conclusion, and even if some correlations between Doppler variability and the activity indices remain, the variability of activity indices seems purely random and, therefore, a Doppler model with only stochastic noise terms (white and/or red noise) should provide a sufficient description of the data. The absence of signals in the indices supports the interpretation of any periodicity in the Doppler data as a planet candidate.

\section{Bayesian analysis of GJ 163 velocities}\label{sec:gj163_analysis}

We analysed the 55 HARPS-TERRA velocities using posterior samplings and by calculating estimates for model probabilities as in e.g. \citet{tuomi2012}, \citet{tuomi2012b}, and \citet{tuomi2012c}. We report the solutions by using maximum \emph{a posteriori} (MAP) estimates and 99\% Bayesian credibility sets (BCSs). In the following subsections, we explore the consequences of analysing the velocities with different noise models, different prior densities, and models containing different numbers of Keplerian signals.

\begin{table*}
\center
\caption{HARPS-TERRA (RV$_{TERRA}$) and HARPS-CCF (RV$_{CCF}$) radial velocities of GJ 163 and the corresponding activity-indices.}\label{tab:rvs}
\begin{tabular}{lcccccccc}
\hline \hline
JD             &    RV$_{TERRA}$   &$\sigma_{TERRA}$ & RV$_{CCF}$ & $\sigma_{CCF}$ & BIS    & FWHM     & S-index    & $\sigma_S$ \\

[days]         &   [ms$^{-1}$] &  [ms$^{-1}$]  &  [ms$^{-1}$]    & [ms$^{-1}$]         & [ms$^{-1}$]  & [ms$^{-1}$]    &        &        \\
\hline
2452942.80391  &   7.44  &   1.19  &     --	 &   --	  &     --     &  --       &   0.836    &  0.015  \\				
2452991.72575  &  -1.60  &   1.57  &   0.489   &  2.88    &  -12.96    &  3005.33  &   0.443	&  0.014  \\
2452998.64588  &  -7.04  &   1.28  &  -6.782   &  2.82    &   -3.35    &  3005.86  &   0.561	&  0.014  \\
2452999.69181  &  -8.01  &   1.60  &  -5.524   &  3.42    &  -14.82    &  3029.92  &   0.536	&  0.014  \\
2453000.59407  &  -1.69  &   0.69  &   1.692   &  2.08    &   -7.25    &  3012.51  &   0.566	&  0.012  \\
2453000.72207  &   0.00  &   1.09  &   0.000   &  2.67    &   -4.97    &  3020.29  &   0.583	&  0.014  \\
2453002.61700  &   6.53  &   1.40  &   2.926   &  3.50    &   -2.19    &  3012.15  &   0.488	&  0.015  \\
2453007.60424  &  -7.78  &   0.67  &  -6.155   &  0.71    &   -8.50    &  3010.96  &   0.617	&  0.007  \\
2453788.54460  &  16.53  &   1.10  &  17.023   &  1.66    &   -6.33    &  3014.03  &   0.482	&  0.010  \\
2453980.88027  &   1.35  &   1.27  &   6.536   &  2.04    &   -9.57    &  3017.94  &   0.589	&  0.012  \\
2453989.89449  &   8.40  &   1.69  &   9.316   &  3.20    &   -4.13    &  3021.51  &   0.636	&  0.015  \\
2454384.86524  &   6.65  &   1.75  &   9.882   &  3.36    &   -2.62    &  3023.88  &   0.736	&  0.016  \\
2454437.71873  &   3.07  &   1.23  &   3.333   &  1.92    &   -3.47    &  3015.69  &   0.606	&  0.012  \\
2454478.65069  &  14.38  &   1.11  &  15.609   &  1.88    &   -8.67    &  3015.50  &   0.613	&  0.011  \\
2454487.57309  &   7.94  &   1.11  &  10.979   &  2.00    &  -11.29    &  3019.68  &   0.633	&  0.012  \\
2454520.60936  &   5.66  &   1.25  &   6.645   &  2.24    &   -7.05    &  3014.15  &   0.764	&  0.014  \\
2454719.89611  &   0.02  &   1.53  &  -3.382   &  2.56    &   -5.21    &  3025.31  &   0.726	&  0.014  \\
2454731.88860  &  -3.10  &   0.94  &  -2.063   &  1.29    &   -6.74    &  3018.23  &   0.633	&  0.009  \\
2454733.81637  &  -4.57  &   1.06  &  -5.737   &  1.41    &   -1.67    &  3020.03  &   0.590	&  0.009  \\
2454751.81085  &  -3.68  &   2.30  &  -9.985   &  3.54    &  -10.50    &  3028.71  &   0.725	&  0.017  \\
2454804.71968  &  -6.88  &   2.68  &  -2.563   &  4.81    &  -20.19    &  3020.77  &   0.807	&  0.028  \\
2454812.61600  &  -0.21  &   1.40  &  -3.919   &  2.17    &   -8.11    &  3028.12  &   0.584	&  0.012  \\
2454825.56635  &  -5.61  &   0.83  &  -3.016   &  1.25    &   -8.70    &  3019.85  &   0.587	&  0.009  \\
2454826.57892  &  -8.83  &   0.81  &  -6.689   &  1.46    &   -7.64    &  3013.95  &   0.574	&  0.009  \\
2454827.60800  & -11.72  &   1.05  & -12.375   &  1.41    &   -3.80    &  3017.42  &   0.553	&  0.009  \\
2454828.62017  & -12.26  &   1.19  & -11.045   &  1.93    &   -8.74    &  3018.72  &   0.542	&  0.011  \\
2454829.58609  &  -8.85  &   1.12  &  -9.272   &  2.14    &   -4.33    &  3020.46  &   0.513	&  0.012  \\
2454830.61306  &  -6.46  &   1.25  &  -4.387   &  2.10    &   -9.18    &  3025.29  &   0.505	&  0.011  \\
2454831.60750  &  -3.66  &   1.03  &  -0.182   &  1.54    &   -6.70    &  3024.51  &   0.589	&  0.009  \\
2454832.61838  &   2.75  &   1.56  &   6.009   &  2.79    &  -12.54    &  3022.82  &   0.545	&  0.013  \\
2454833.60932  &   2.06  &   1.30  &   1.174   &  2.37    &  -11.77    &  3030.98  &   0.656	&  0.012  \\
2454834.65245  &   2.06  &   0.94  &   5.226   &  1.45    &  -10.34    &  3022.10  &   0.720	&  0.010  \\
2454840.62729  &   4.66  &   0.99  &   4.379   &  1.73    &   -9.27    &  3022.76  &   0.657	&  0.011  \\
2454848.54360  &  -0.98  &   1.29  &  -0.022   &  2.22    &   -9.43    &  3026.88  &   0.615	&  0.012  \\
2454849.55288  &  -0.15  &   0.71  &   1.103   &  1.20    &   -9.88    &  3020.24  &   0.639	&  0.009  \\
2454850.56610  &   1.74  &   1.23  &   1.981   &  1.90    &   -8.42    &  3023.12  &   0.713	&  0.012  \\
2454851.61088  &  -1.96  &   0.73  &   1.326   &  1.37    &  -10.1     &  3016.54  &   0.700	&  0.010  \\
2454852.61939  &  -3.93  &   0.94  &  -4.340   &  1.59    &   -9.80    &  3016.81  &   0.709	&  0.011  \\
2454854.60026  &  -6.26  &   0.89  &  -4.183   &  1.82    &  -15.0     &  3023.30  &   0.741	&  0.012  \\
2454871.56247  &  -4.98  &   1.13  &  -4.297   &  1.72    &   -2.62    &  3028.80  &   0.613	&  0.010  \\
2454878.55525  &   0.71  &   0.97  &   2.795   &  1.95    &   -4.88    &  3020.09  &   0.568	&  0.012  \\
2454879.57723  &  -2.71  &   1.35  &  -3.872   &  2.03    &   -0.27    &  3013.55  &   0.680	&  0.013  \\
2454880.55542  &  -3.42  &   0.91  &  -3.103   &  1.56    &   -7.65    &  3015.34  &   0.640	&  0.011  \\
2454881.61112  &  -0.21  &   1.25  &  -1.159   &  2.41    &   -6.82    &  3014.42  &   0.477	&  0.013  \\
2454882.55717  &   3.75  &   1.21  &   5.711   &  1.97    &   -7.28    &  3013.30  &   0.677	&  0.013  \\
2454883.54916  &   9.15  &   1.12  &  10.195   &  1.83    &   -6.94    &  3013.92  &   0.603	&  0.012  \\
2454884.57656  &  12.61  &   1.70  &  11.976   &  2.33    &   -7.65    &  3014.88  &   0.447	&  0.012  \\
2454885.52811  &  13.88  &   1.17  &  14.523   &  1.59    &  -10.91    &  3015.93  &   0.391	&  0.008  \\
2454886.53336  &   6.00  &   1.27  &   7.112   &  2.22    &  -12.88    &  3001.02  &   0.570	&  0.014  \\
2454914.55044  &   0.78  &   1.91  &   2.875   &  3.27    &   -9.24    &  3018.98  &   0.746	&  0.019  \\
2454916.50856  &   5.97  &   1.06  &      --   &  --      &      --    &      --   &   0.685	&  0.012  \\
2454918.50265  &  12.16  &   0.87  &      --   &  --      &      --    &      --   &   0.569	&  0.010  \\
2454920.52842  &   9.70  &   0.97  &      --   &  --      &      --    &      --   &   0.640	&  0.010  \\
2455423.83105  &  -1.59  &   2.19  &   3.299   &  3.77    &  -13.93    &  3019.61  &   0.490	&  0.017  \\
2455437.82442  &   2.81  &   0.99  &   3.615   &  1.52    &   -7.04    &  3014.29  &   0.700	&  0.013  \\
\hline \hline
\end{tabular}
\end{table*}

\subsection{Reference priors, white noise}

We started the analyses of GJ 163 velocities by using the reference priors as defined in Table \ref{tab:ref_priors}. We used the common white noise model and assumed that any planets orbiting the star do not interact with one another on the time-scale of the observational baseline.

With these assumptions, we estimated the Bayesian evidence for a reference model with $k=0$ and searched for periodicities in the data by using a model with $k=1$. Our Markov chains converged to a clear signal at a period of 8.63 days. Accounting for this signal increased the model probability to a value 7.3$\times 10^{5}$ times greater than for the model with $k=0$.

The two-Keplerian model increased the model probabilities considerably by a factor of 8.0$\times 10^{10}$, implying the presence of another periodicity in the data. The second periodicity was found at 225 days and we could constrain this emerging signal according to our detection criteria shown in \citet{tuomi2012}. Samplings of the parameter space of a three-Keplerian model revealed a third significant periodicity at 25.6 days and the corresponding model was found to have a posterior probability of roughly 4.1$\times 10^{8}$ times that of the model with $k = 2$. However, the MAP periods corresponding to this model ($k = 3$) were found to be 8.63, 25.6, and 567 days and although we found a local maximum in the period space at 225 days, it does not correspond to a global solution of the three-Keplerian model anymore. We searched for additional periodicities in the data as well, but our Markov chains did not converge to a well-constrained solution for the four-Keplerian model.

\begin{figure}
\center
\includegraphics[angle=-90,width=0.45\textwidth]{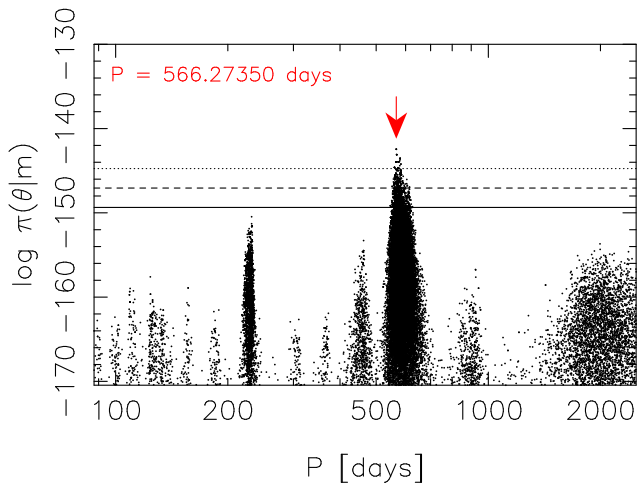}
\caption{Log-posterior density as a function of the longest periodicity in the GJ 163 data based on temperate sampling. The red arrow indicates the highest value in the sample roughly corresponding to the MAP estimate and the horizontal lines show the probability thresholds 10\% (dotted line), 1\% (dashed line), and 0.1\% (solid line) of the maximum.}\label{fig:temperate}
\end{figure}

To demonstrate the robustness of our solution with $k=3$, we performed temperate samplings of the parameter space of the model such that instead of obtaining samples from the posterior density ($\pi$) as such, we used $\pi^{\beta}$ as a posterior density, where $\beta$ is a parameter in the interval [0,1] describing the ``temperature'' of the sampling. In particular, we chose $\beta = 0.5$, and plotted the obtained Markov chain in Fig. \ref{fig:temperate}. In this figure, we show the posterior density corresponding to the temperate sampling as a function of the longest periodicity. It can be seen that the 567-day period corresponds to the global maximum but that there are local maxima as well at periods of roughly 220, 450, and 2000 days. However, none of the local maxima exceeds 0.1\% level of the global maximum (solid horizontal line in Fig. \ref{fig:temperate}), which indicates that the 567-day periodicity is clearly the preferred solution of this model.

\subsection{Reference priors, white and correlated noise}

\begin{figure}
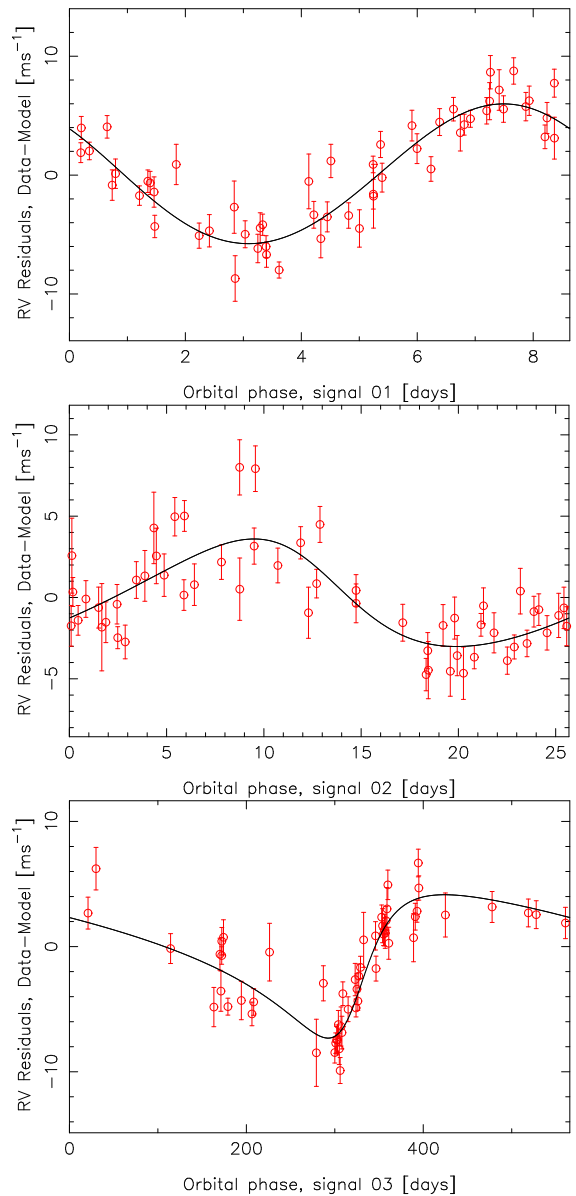

\center
\includegraphics[angle=-90,width=0.40\textwidth]{rvdist03_scresidc_rv_GJ163_1.ps}

\includegraphics[angle=-90,width=0.40\textwidth]{rvdist03_scresidc_rv_GJ163_2.ps}

\includegraphics[angle=-90,width=0.40\textwidth]{rvdist03_scresidc_rv_GJ163_3.ps}
\caption{Phase-folded signals of the three-Keplerian model with the other three signals subtracted from each panel.}\label{fig:signals_3p}
\end{figure}

As pointed out by \citet{baluev2012}, \citet{tuomi2012d}, and \citet{tuomi2012e}, RV variations might appear due to another noise component that can be described as having a red colour. This noise component can be accounted for by adding a correlation term in the model of the $i$th measurement with the previous ones. We applied a moving average (MA) noise model \citep{tuomi2012c,tuomi2012d,tuomi2012e} and observed that even with a $k=0$ model, the Bayesian evidences implied that the model including this type of red noise was considerably better than the one with only white noise. We have listed the resulting log-Bayesian evidences ($\ln P(m)$) of pure white noise and moving average models with $k=0, ..., 3$ in Table \ref{tab:reference_prior_probs} and plotted the corresponding phase-folded Keplerian signals in Fig. \ref{fig:signals_3p}.

\begin{table}
\center
\caption{Log-Bayesian evidences of models with $k=0, ..., 3$ and pure white noise or an MA noise model. The evidences are estimates assuming the reference priors.}\label{tab:reference_prior_probs}
\begin{tabular}{lcc}
\hline \hline
$k$ & $\ln P(m)$ & $\ln P(m)$ \\
& White & MA \\
\hline
0 & -185.6 & -168.9 \\
1 & -171.4 & -153.4 \\
2 & -145.6 & -136.9 \\
3 & -125.0 & -127.1 \\
\hline \hline
\end{tabular}
\end{table}

A noteworthy feature in Table \ref{tab:reference_prior_probs} is that the MAP solution of the MA model with $k=2$ differs from that of the pure white noise model by having a second periodic signal at a period of 25.6 days instead of 225 days. Therefore, it is likely that the apparent periodicity at 225 days can be interpreted as being caused by a combination of yearly aliasing with the 567-day signal (because 1/365 $\approx$ 1/225 $-$ 1/567) and/or noise correlations rather than as a genuine Keplerian signal. Indeed, the two-Keplerian model with an MA component has a much higher posterior probability than that with pure white noise. This implies that, once two Keplerian signals are included in the model, the RV noise of subsequent epochs is not independent for $k=2$. For $k=3$ the situation is reversed and the white noise model is in fact better than the MA model. We believe this is the case because the MA model is slightly overparameterised due to the rather low number of measurements (55) with respect to the number of free parameters (18) and thus penalised by Occam's razor -- the MAP estimate of the correlation parameter $\phi$ is clearly positive (0.83) but negligible correlations are also allowed (with a 99\% BCS of [0.01, 1]), which implies that the MA component is not necessary for the three-Keplerian model. As was the case with the white noise model, we could not find a significant fourth periodic signal in the data using posterior samplings of a four-Keplerian model and MA noise.

Using the reference prior, we would conclude that there are confidently three Keplerian signals in the data. We have listed the best solution in Table \ref{tab:reference_solution_3p}. We note that this solution corresponds to an outer planet with an eccentric orbit ($e=0.46$), which casts doubt on the stability of the corresponding system in long term. For this reason, we test eccentricity priors that penalise high eccentricities more than the reference prior.

\begin{table*}
\center
\caption{MAP estimates and the corresponding 99\% BCSs of the parameters of the three-Keplerian model for the reference priors and a pure white noise model.}\label{tab:reference_solution_3p}
\begin{tabular}{lccc}
\hline \hline
Parameter & GJ 163 b & GJ 163 c & GJ 163 d \\
\hline
$P$ [d] & 8.6313 [8.6259, 8.6363] & 25.662 [25.513, 25.751] & 567 [544, 587] \\
$e$ & 0.03 [0, 0.23] & 0.08 [0, 0.48] & 0.46 [0.29, 0.67] \\
$K$ [ms$^{-1}$] & 5.55 [4.17, 6.78] & 3.30 [1.98, 4.62] & 5.09 [3.39, 7.19] \\
$\omega$ [rad] & 0.1 [0, 2$\pi$] & 0.5 [0, 2$\pi$] & 4.03 [3.44, 4.70] \\
$M_{0}$ [rad] & 4.3 [0, 2$\pi$] & 3.0 [0, 2$\pi$] & 3.3 [1.4, 5.7] \\
$m_{p} \sin i$ [M$_{\oplus}$] & 9.6 [6.8, 12.7] & 8.2 [4.6, 11.7] & 35 [21, 52] \\
$a$ [AU] & 0.061 [0.054, 0.067] & 0.125 [0.113, 0.137] & 0.98 [0.87, 1.08] \\
\hline
$\gamma$ [ms$^{-1}$] & 1.54 [0.14, 2.94] \\
$\sigma_{J}$ [ms$^{-1}$] & 1.52 [0.76, 2.52] \\
\hline \hline
\end{tabular}
\end{table*}

\subsection{Alternative priors: $\sigma_e$=0.1 and 0.2, white noise}

Using a more restrictive forms for the prior density of eccentricity, i.e. priors with $\sigma_{e} = 0.2$ and $\sigma_{e} = 0.1$ (see Table \ref{tab:ref_priors}), we repeated the analyses of the GJ 163 velocities with the pure white noise model.

According to our posterior samplings and estimations of Bayesian evidences, when using models with $k=0, ..., 5$ and assuming that all the excess noise in the data is white, the model probabilities increased as $k$ approached four but we could not estimate the Bayesian evidences reliably for $k=5$ because we could not spot a fifth periodicity in the data reliably according to our criteria. However, we observed a broad probability maximum in the period space at a period of roughly 1500 days with a low MAP amplitude of 2.7 ms$^{-1}$. Specifically, a fifth periodicity could not be constrained from above and below in the period space and we therefore conclude that this prior choice and white noise model favours the existence of four periodic signals in the data. We obtained the same qualitative result for both eccentricity priors.

We have listed the log-Bayesian evidences of models with $k=0, ..., 4$ in Table \ref{tab:ecc_prior_probs}. According to these results, the eccentricity prior with $\sigma_{e} = 0.2$ is much better prior model for the data because it allows reasonably eccentric solutions that are favoured by the data for $k=2, 3$, and 4. The reason is that the eccentricity of the second strongest signal with a period of 567 days has MAP estimates of approximately 0.45 for models $k=2$ and $k=3$, respectively. These eccentricities are penalised much more severely with the prior with $\sigma_{e} = 0.1$ which results in decreased Bayesian evidences because the prior actually conflicts with the likelihood function that favours higher eccentricities.

\begin{table}
\center
\caption{Log-Bayesian evidences of models with $k=0, ..., 4$ and pure white noise or an MA noise model. Estimates are shown for eccentricity priors with $\sigma_{e} = 0.1$ (E1) and $\sigma_{e} = 0.2$ (E2).}\label{tab:ecc_prior_probs}
\begin{tabular}{lcccc}
\hline \hline
$k$ & $\ln P(m)$ & $\ln P(m)$ & $\ln P(m)$ & $\ln P(m)$ \\
 & White E1 & White E2 & MA E1 & MA E2 \\
\hline
0 & -185.6 & -185.6 & -168.9 & -168.9 \\
1 & -171.3 & -171.5 & -153.6 & -153.5 \\
2 & -154.4 & -147.0 & -136.7 & -137.1 \\
3 & -131.7 & -124.2 & -129.7 & -127.5 \\
4 & -125.3 & -119.9 & -- & -- \\
\hline \hline
\end{tabular}
\end{table}

\begin{figure}
\center
\includegraphics[angle=-90,width=0.40\textwidth]{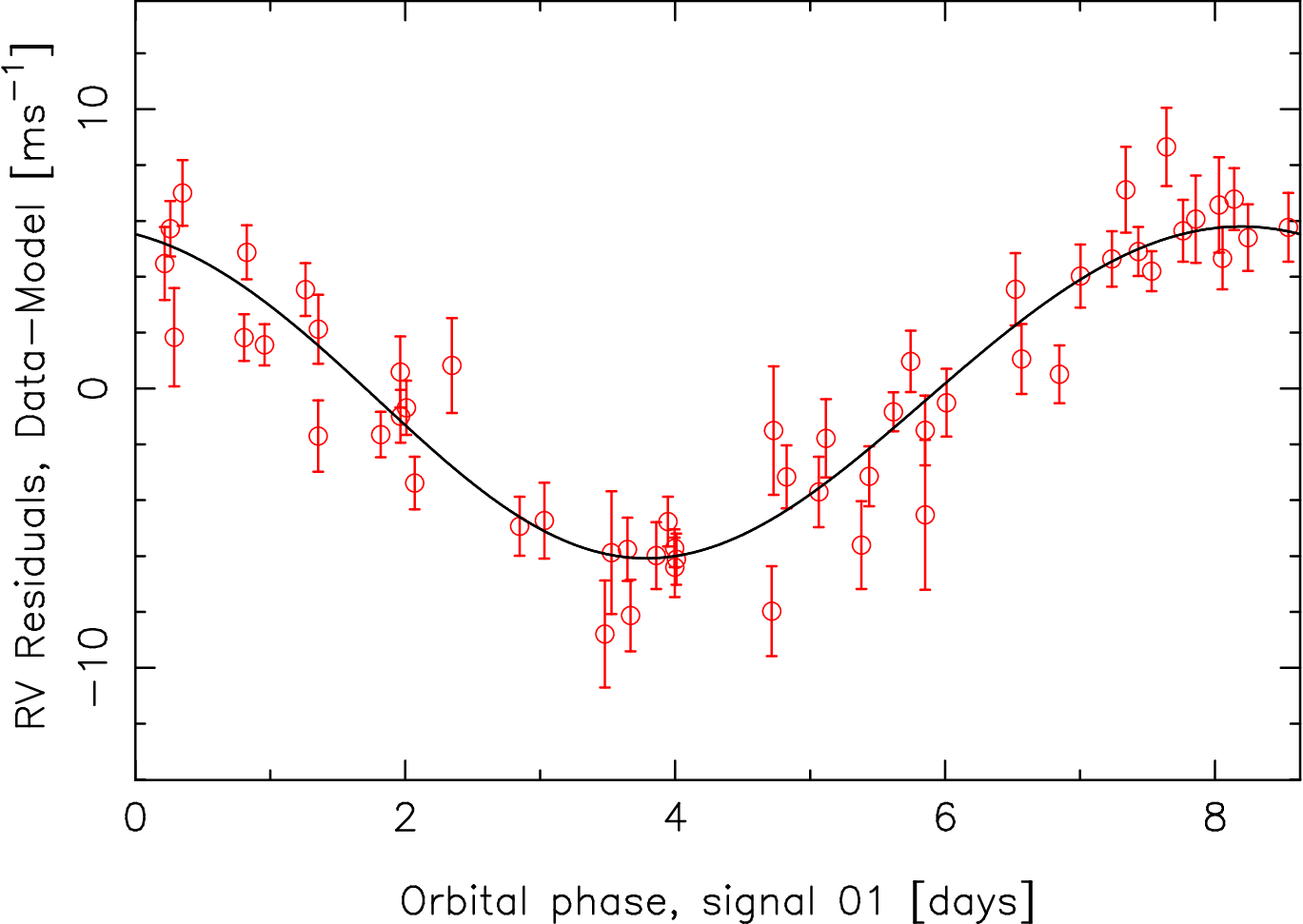}

\includegraphics[angle=-90,width=0.40\textwidth]{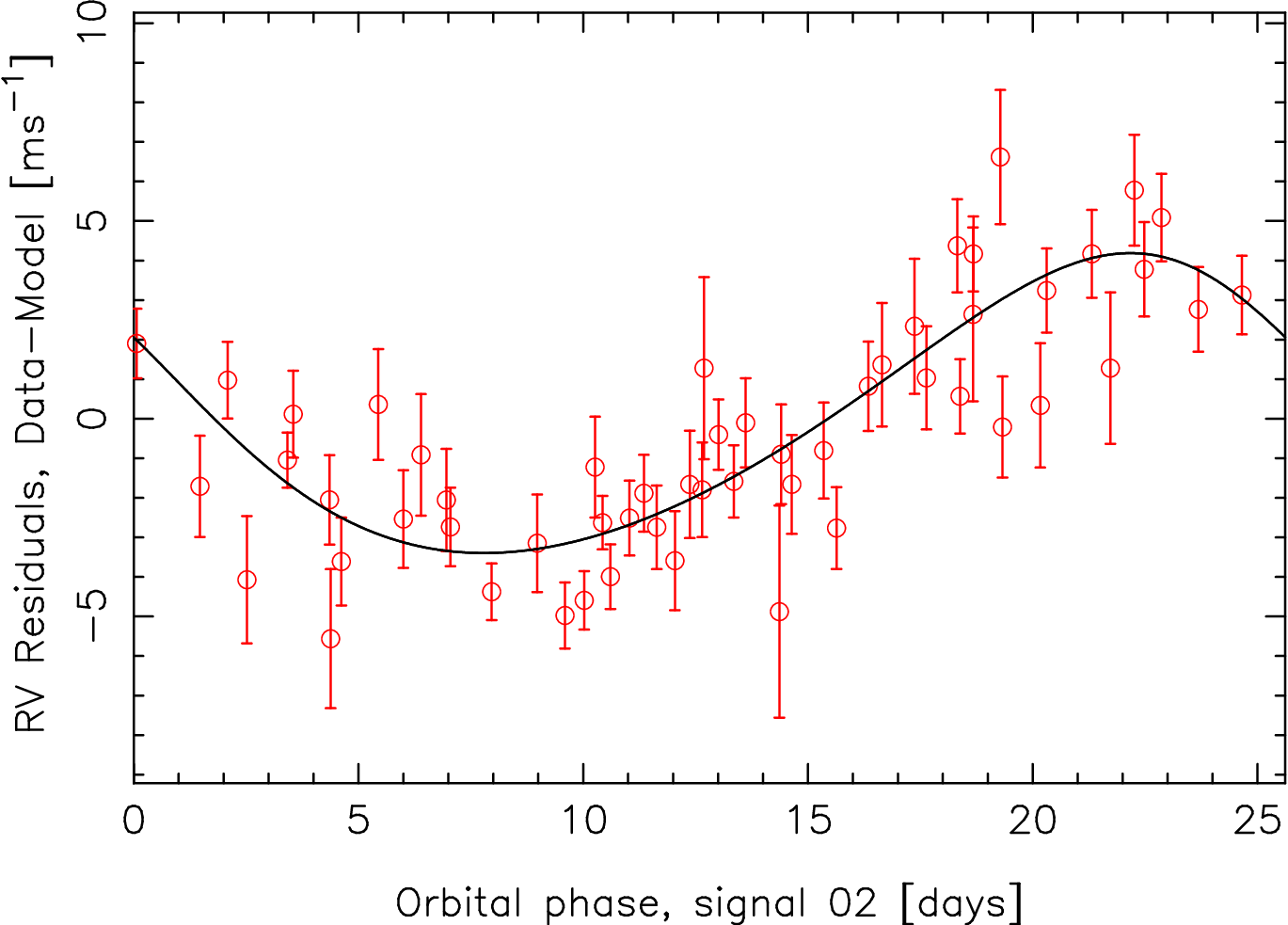}

\includegraphics[angle=-90,width=0.40\textwidth]{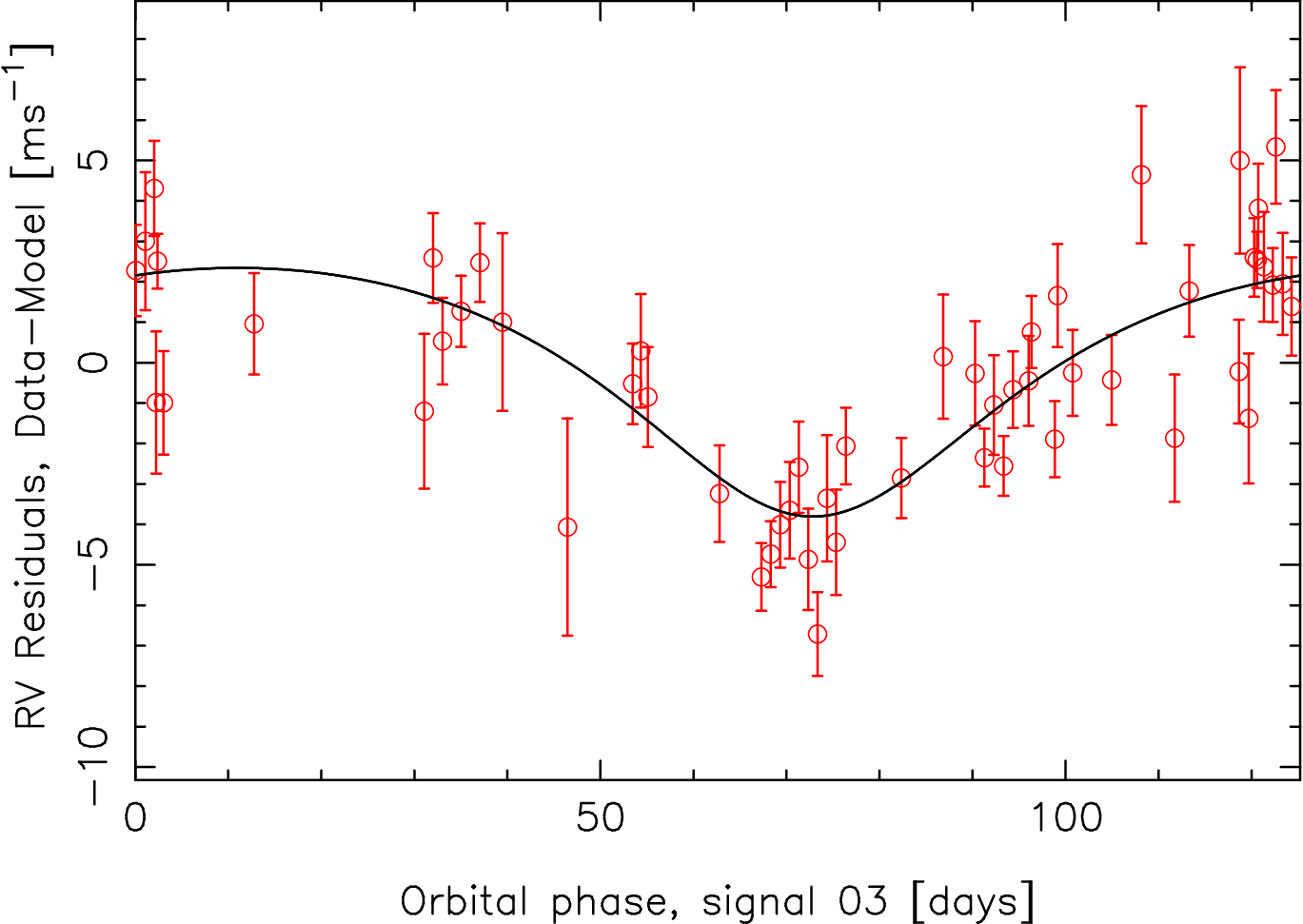}

\includegraphics[angle=-90,width=0.40\textwidth]{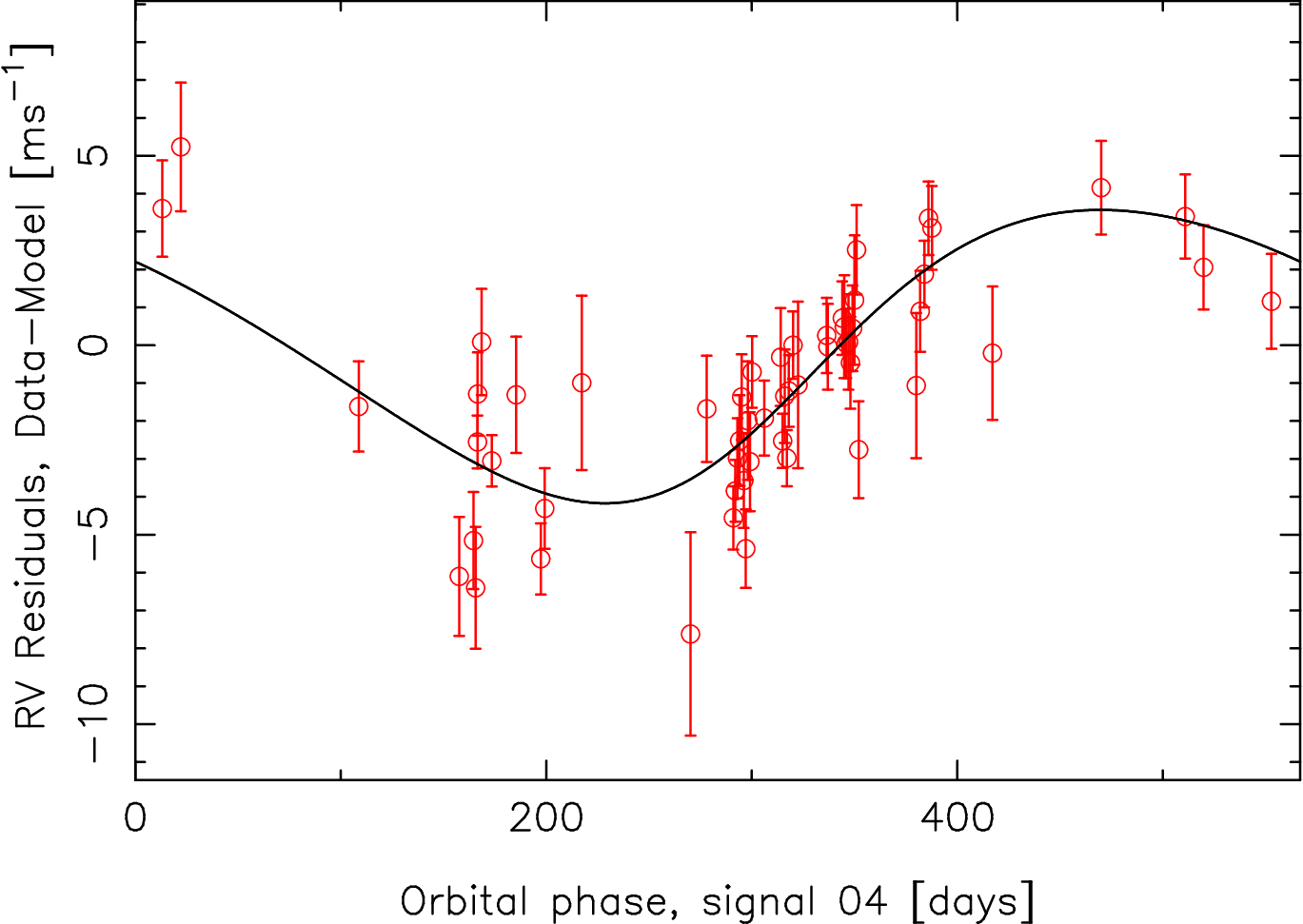}
\caption{Phase-folded signals of the four-Keplerian model with the other three signals subtracted from each panel.}\label{fig:phase_orbits}
\end{figure}

\begin{figure}
\center
\includegraphics[angle=-90,width=0.40\textwidth]{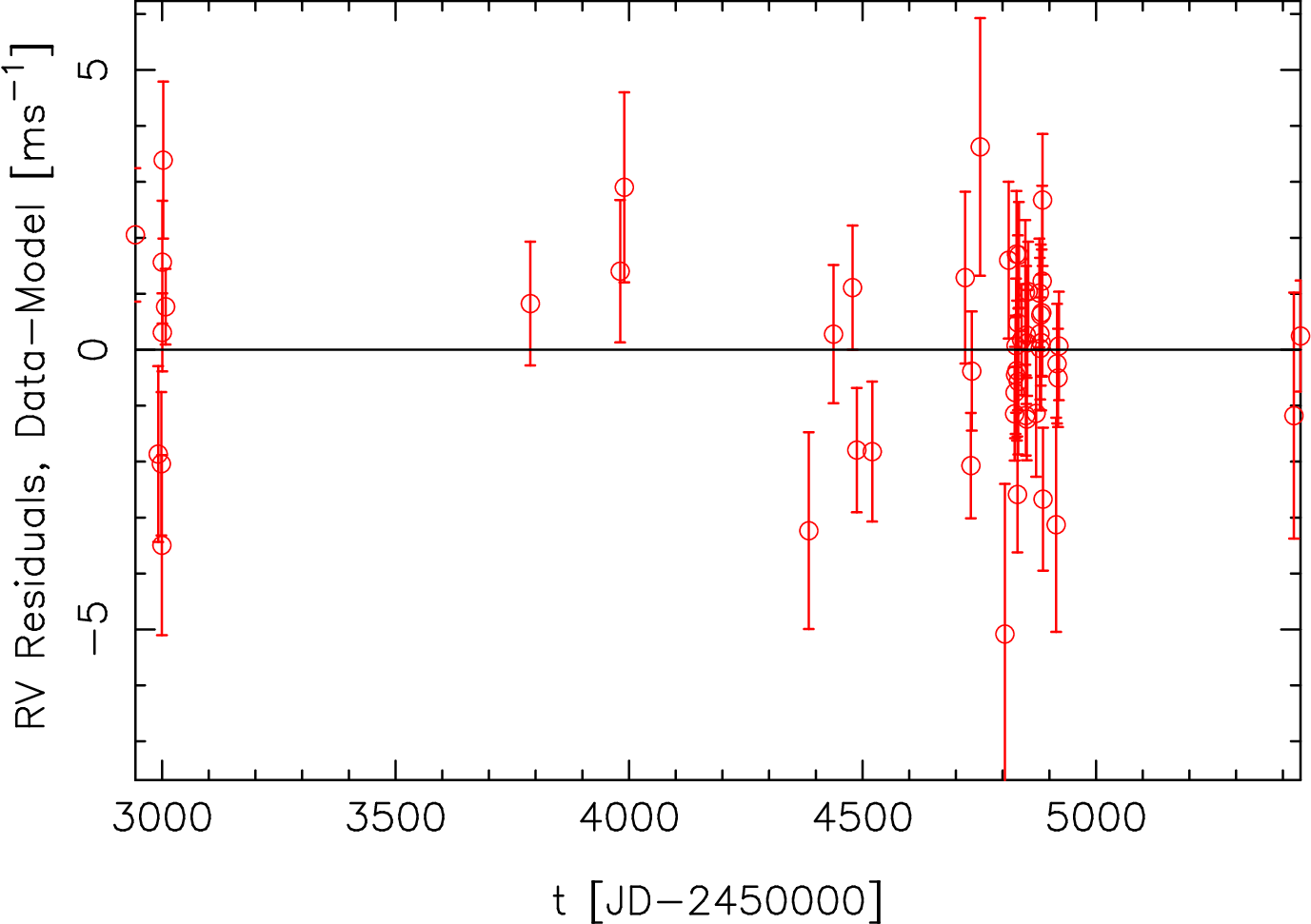}
\caption{Residuals of the four-Keplerian orbital solution shown in Fig. \ref{fig:phase_orbits}.}\label{fig:residuals}
\end{figure}

For a white noise model with $k=4$ the favoured eccentricities of the signal at a period of 567 days and a fourth signal emerging at 125 days are not very high but they still have MAP estimates of roughly 0.2. Yet, eccentricities as high as 0.5 cannot be ruled out either, which is penalised by the eccentricity prior with $\sigma_{e} = 0.1$ and results in a decreased Bayesian evidence for this more restrictive prior (Table \ref{tab:ecc_prior_probs}). Therefore, assuming close-to-circular orbits and that the RV noise is white, we would conclude that the is evidence in favour of four signals in the GJ 163 data. We plotted the phase-folded signals of the four-Keplerian model in Fig. \ref{fig:phase_orbits} and the remaining RV residuals after subtracting the MAP signals in Fig. \ref{fig:residuals} to demonstrate that our model also reproduces the data visually.

\subsection{Alternative priors: $\sigma_e$=0.1 and 0.2, white and correlated noise}

The situation changes considerably when taking into account the possible correlations in the data using the MA model. While there is not much difference in the performance of these two eccentricity priors (Table \ref{tab:ecc_prior_probs}), a fourth periodicity cannot be found according to our detection criteria.

\begin{figure}
\center
\includegraphics[angle=-90,width=0.4\textwidth]{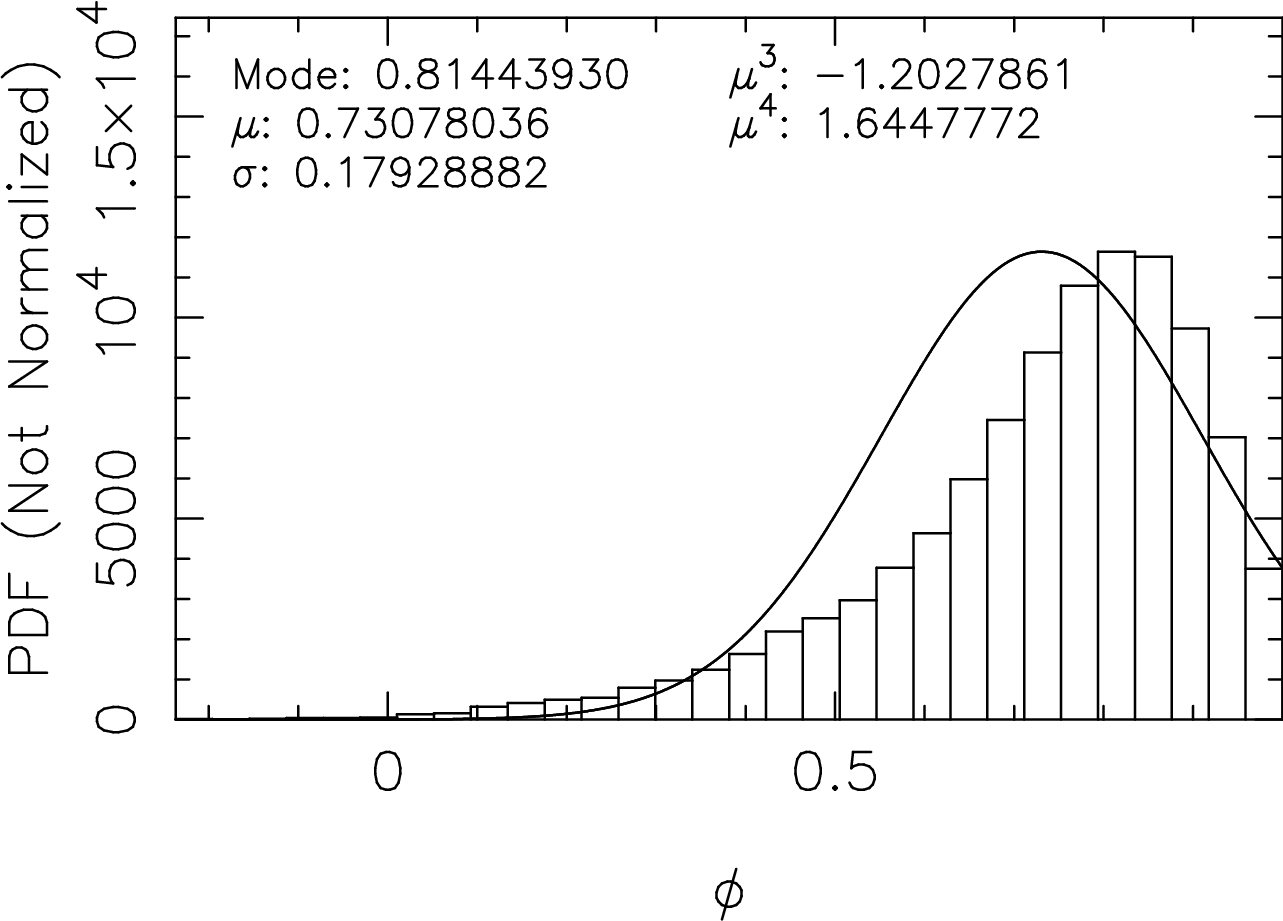}
\caption{Distribution of the correlation parameter for an MA noise model and $k=3$ with a hyperparameter $\sigma_{e} = 0.2$.}\label{fig:correlation_dist}
\end{figure}

In absolute terms, the model that has the highest posterior probability given the GJ 163 data is the white noise model with $\sigma_{e} = 0.2$. While the MA model is much better for $k=0, 1$, and 2 with any of the eccentricity priors, the pure white noise model is favoured by the data for $k=3$ and is even better for $k=4$ (Table \ref{tab:ecc_prior_probs}). The likely reason for this result is that the MA parameter $\phi$ is actually consistent with zero for $k=3$ (Fig. \ref{fig:correlation_dist}), despite that its MAP estimate is 0.83, which makes the model overparameterised and decreases its Bayesian evidence estimate. 

\begin{table*}
\center
\caption{MAP estimates and the corresponding 99\% BCSs of the parameters of the model with the greatest posterior probability containing four Keplerian signals and pure white noise and a hyperparameter $\sigma_{e} = 0.2$.}\label{tab:best_solution_4p}
\begin{tabular}{lcccc}
\hline \hline
Parameter & GJ 163 b & GJ 163 c & (GJ 163 e) & GJ 163 d \\
\hline
$P$ [d] & 8.6312 [8.6267, 8.6350] & 25.632 [25.557, 25.715] & 125.0 [123.3, 128.0] & 572 [531, 603] \\
$e$ & 0.02 [0, 0.18] & 0.01 [0, 0.38] & 0.32 [0, 0.55] & 0.27 [0, 0.54] \\
$K$ [ms$^{-1}$] & 5.87 [4.55, 7.05] & 3.54 [2.20, 4.75] & 3.38 [1.03, 4.92] & 3.76 [1.88, 5.87] \\
$\omega$ [rad] & 5.0 [0, 2$\pi$] & 2.7 [0, 2$\pi$] & 2.7 [0, 2$\pi$] & 2.8 [0, 2$\pi$] \\
$M_{0}$ [rad] & 4.3 [0, 2$\pi$] & 1.6 [0, 2$\pi$] & 3.1 [0, 2$\pi$] & 0.7 [0, 2$\pi$] \\
$m_{p} \sin i$ [M$_{\oplus}$] & 9.9 [7.2, 13.3] & 8.7 [5.3, 12.3] & 14.0 [4.5, 21.6] & 27 [11, 43] \\
$a$ [AU] & 0.061 [0.054, 0.067] & 0.126 [0.112, 0.138] & 0.361 [0.325, 0.394] & 1.00 [0.87, 1.10] \\
\hline
$\gamma$ [ms$^{-1}$] & 1.58 [-0.25, 2.98] \\
$\sigma_{J}$ [ms$^{-1}$] & 1.35 [0.46, 2.33] \\
\hline \hline
\end{tabular}
\end{table*}

\begin{figure*}
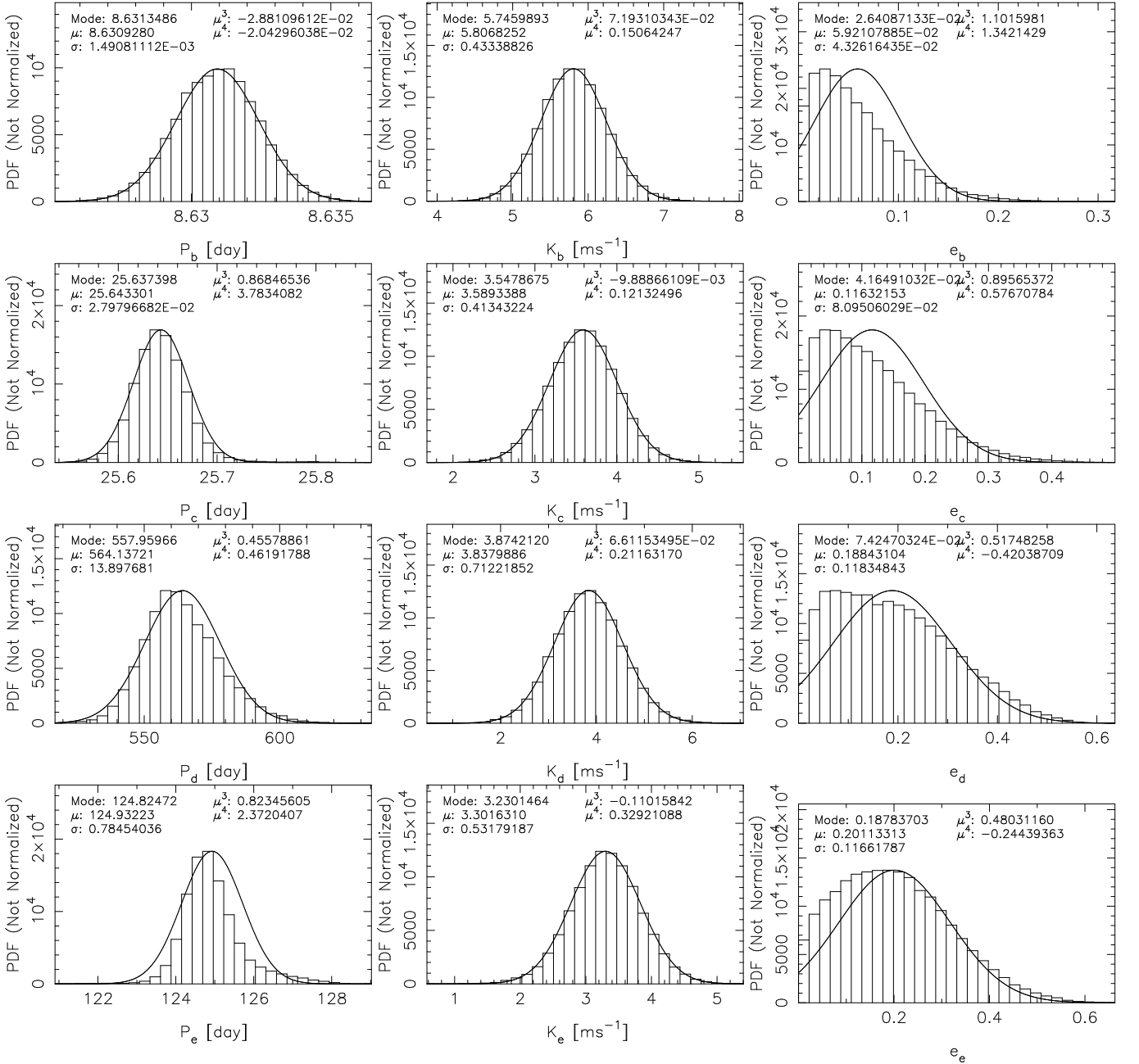

\center
\includegraphics[angle=-90,width=0.32\textwidth]{rvdist04_rv_GJ163_dist_Pb.ps}
\includegraphics[angle=-90,width=0.32\textwidth]{rvdist04_rv_GJ163_dist_Kb.ps}
\includegraphics[angle=-90,width=0.32\textwidth]{rvdist04_rv_GJ163_dist_eb.ps}

\includegraphics[angle=-90,width=0.32\textwidth]{rvdist04_rv_GJ163_dist_Pc.ps}
\includegraphics[angle=-90,width=0.32\textwidth]{rvdist04_rv_GJ163_dist_Kc.ps}
\includegraphics[angle=-90,width=0.32\textwidth]{rvdist04_rv_GJ163_dist_ec.ps}

\includegraphics[angle=-90,width=0.32\textwidth]{rvdist04_rv_GJ163_dist_Pd.ps}
\includegraphics[angle=-90,width=0.32\textwidth]{rvdist04_rv_GJ163_dist_Kd.ps}
\includegraphics[angle=-90,width=0.32\textwidth]{rvdist04_rv_GJ163_dist_ed.ps}

\includegraphics[angle=-90,width=0.32\textwidth]{rvdist04_rv_GJ163_dist_Pe.ps}
\includegraphics[angle=-90,width=0.32\textwidth]{rvdist04_rv_GJ163_dist_Ke.ps}
\includegraphics[angle=-90,width=0.32\textwidth]{rvdist04_rv_GJ163_dist_ee.ps}
\caption{Marginal distributions of the orbital periods ($P_{x}$), eccentricities ($e_{x}$), and RV amplitudes ($K_{x}$) of the four-Keplerian solution in Table \ref{tab:best_solution_4p}. The solid curves are Gaussian densities with the same mean and variance as the marginal distributions.}\label{fig:densities}
\end{figure*}

We report the best solution corresponding to the white noise model and four Keplerian signals in Table \ref{tab:best_solution_4p} and Fig. \ref{fig:densities}. The corresponding probabilities of this model with the \emph{moderate} eccentricity prior with $\sigma_{e}=0.2$ are 4.8$\times 10^{-28}$, 3.2$\times 10^{-22}$, 7.1$\times 10^{-12}$,  0.026, and 0.974 for $k=0, 1, 2, 3$, and 4, respectively. We note that the four-Keplerian model does not have a posterior probability that exceeds the detection threshold (by being at least 150 times greater than that of the three-Keplerian model). While the probability of the model with $k=4$ is \emph{only} $\sim$ 38 times greater than that of the model with $k=3$, the other two detection conditions are satisfied -- all periods are well-constrained and the amplitudes are statistically different from zero, as can be seen in Fig. \ref{fig:densities}. Therefore, we present a tentative four-Keplerian solution as the preferred one.

We note that using a proper informative prior for the jitter parameter ($\sigma_{J}$) did not change the results significantly for any of the statistical models. The jitter prior with  $\mu_{\sigma} = 1$ ms$^{-1}$ and $\sigma_{\sigma} = $1 ms$^{-1}$ penalised models with $k=0$ and $k=1$ slightly but does not have a significant effect on neither the Bayesian evidences nor the obtained posterior densities. We believe this happens because the jitter prior does not represent the data very well for $k=0$ and 1 because the two corresponding models contain at least two or three signals that increase the variability of the data the models interpret as noise. This can be seen as additional support for the existence of more than one signals in the data. However, conclusions can only be based on the Bayesian evidences and the corresponding model probabilities, unless there was a very strong \emph{a priori} reason to believe that the excess jitter cannot exceed 1-2 ms$^{-1}$ level considerably, which is not the case here.

\subsection{Period prior comparison}

Finally, we tested if the results were affected significantly by the chosen period prior, i.e. whether the prior was constructed by assuming a uniform density in the period space, orbital frequency space, or log-period space. We tested this using the three-Keplerian model and an MA noise model and calculated the log-Bayesian evidences using each period prior. We obtained log-Bayesian evidences that were within 0.5 from one another, which cannot be considered a significant difference in any model selection problem because it corresponds to probabilities between 0.23 -- 0.45 for the three models. The likely reason is that in the vicinity of the period maxima, all these period priors are roughly constant and constant coefficients do not have an effect on the estimates of Bayesian evidences. Also, according to our tests with different values of $k$, the different period priors did not affect the ability to detect signals in the data in the first place compared to the reference prior that was uniform in log-period.

\section{Dynamical feasibility of the system}

As a final validation of the signals as planet candidates, we performed a dynamical analysis on representative samples of parameters drawn from the posterior densities. According to the integrations we performed by using the Bulirsch-Stoer algorithm \citep{bulirsch1966}, all parameter vectors drawn from the posterior density of the model with $k=3$ (samples compatible with Table \ref{tab:best_solution_4p}) corresponded to stable planetary systems on a time-scale of 10$^{6}$ years or longer. We repeated the same experiment with samples from the posterior density of the model with $k=4$ (represented by Table \ref{tab:best_solution_4p}). Again, all vectors corresponded to stable orbital configurations beyond 10$^6$ years and, therefore, we could not use any dynamical argument to decide which solution was the most favored one from a physical point of view.

\begin{figure}
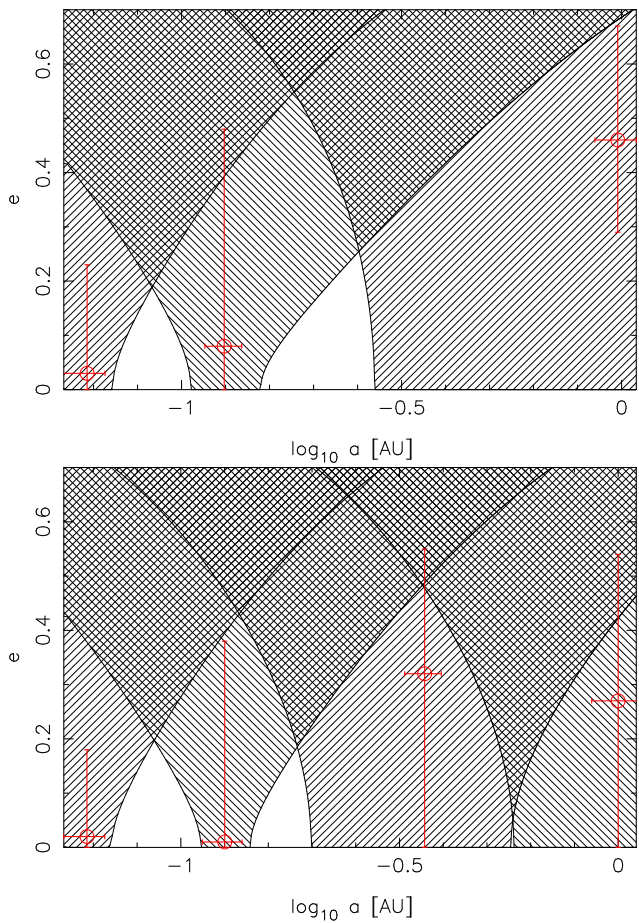

\center
\includegraphics[angle=-90,width=0.45\textwidth]{stabilitybound_GJ163_3.ps}

\includegraphics[angle=-90,width=0.45\textwidth]{stabilitybound_GJ163_4.ps}
\caption{Approximated Lagrange stability boundaries indicating the parameter space around the MAP estimates (red circles) where there are no stable orbits (shaded areas). Three- (top) and four-planet (bottom) solutions.}\label{fig:stability}
\end{figure}

According to our numerical integrations of the orbits, the orital elements remained in bounded areas of the parameter space without corresponding to collisions, orbital crossings, or escapes from the system (beyond 10 AU), in accordance with the Lagrange stability criteria. Following \citet{tuomi2012} and \citet{tuomi2012d}, we demonstrate this by plotting the approximated Lagrange stability boundaries of the solutions in Tables \ref{tab:reference_solution_3p} and \ref{tab:best_solution_4p} in Fig. \ref{fig:stability}. This figure illustrates that our three- and four-planet solutions correspond to planetary systems with sufficient orbital spacing to enable long-term stability.

Finally, we tested the chaotic behaviour of our solutions by using the frequency analysis method of \citet{laskar1993} as applied in e.g. \citet{correia2010} and \citet{tuomi2012c}. According to our results, the relative variation in the mean motions of the planets ($D$-index) is at most $10^{-4}$ for the most eccentric solutions of the three-Keplerian model, indicating that such orbits might show some chaotic behaviour, whereas typically we found this variation to be roughly $10^{-6}$ or even less, which indicates very regular and thus stable motion.

\section{Discussion and conclusions}

Based on our analyses of the GJ 163 velocity data from the HARPS spectra, the absence of any clear periodicity in the activity indices, and the dynamical feasibility of the orbital solutions, we conclude that GJ 163 has at least three planet candidates orbiting it. This conclusion is independent of noise models and prior densities and is therefore very strongly supported by the data. When using a white noise model and slightly more limiting eccentricity priors, we obtained a solution for the four-Keplerian model that was well constrained in the parameter space (Table \ref{tab:best_solution_4p} and Fig. \ref{fig:densities}) but did not exceed the detection threshold of being 150 times more probable than the three-Keplerian model. We interpret this result as suggestive but inconclusive evidence for a fourth planet candidate in the system. Given that two of the three detection conditions of \citet{tuomi2012} are satisfied for this fourth candidate, even a small amount of additional measurements ($\sim$ 20 new observations) might settle the issue in favour or against the fourth candidate. Additional measurements are also strongly encouraged to investigate whether a very broad local probability maximum we observed in the period space at roughly 1500 days corresponds to yet another candidate.

Having presented results based on different models and priors, we discuss briefly how one should proceed in similar analyses of RV data from HARPS and other instruments. First, it is crucial to investigate whether autocorrelation in a suitable timescale can explain some of the variation in the data in comparison to models where this variation is described using a Keplerian model. If autocorrelation -- in a timescale of few dozen days \citep{tuomi2012d,tuomi2012e,baluev2012} -- is a good description of the data and the inclusion of Keplerian signals instead of autocorrelation does not improve the model much, one cannot safely claim that there is evidence for additional planetary signals in the data. However, if the signals satisfy the detection criteria of \citet{tuomi2012} even with an MA model, or a similar model that accounts for the correlated noise, it is likely that the signals are real and possibly of planetary origin if they do not have activity-related counterparts. It is always possible that any given statistical model is not an adequate description of the velocity variations \citep[e.g.][]{tuomi2011b} and, as a consequence, some signals may still be spurious artefacts of poor modelling. Detecting low-mass planets around nearby stars is not only a matter of finding significant signals -- we can only start calling them planetary candidates if these signals are 1) reasonably independent of the exact choice of a noise model (given some good candidate noise models), 2) do not have counterparts in the stellar activity data, and 3) correspond to planetary systems that are physically viable.

The same applies to the choice of priors. We have studied the consequences of assuming different informative priors for the orbital eccentricity. According to our results, the exact choice of this prior does not affect the results much. The only exception in case of GJ 163 velocities is the evidence in favour of a fourth companion when orbital eccentricities are kept \emph{a priori} close to zero.

We also investigated the effect of uninformative prior densities of the period parameter that correspond to uniform priors on different but still plausible parameterisations. For GJ 163 data, the precise choice of the prior did not affect the results in a significant way.

We conclude that the detections of planets in RV data does not seem to be very sensitive to prior choices in practice as long as the chosen priors are not too informative w.r.t. the likelihood functions, i.e. delta-function priors or other very narrow probability densities. As a general rule, we find it is always advisable to repeat the statistical analyses by assuming a few different prior choices to investigate under which hypotheses solutions corresponding to low-amplitude planets are supported by the data (e.g. does one really need to assume close-to-circular orbits to obtain physically viable solutions?).

According to our results \citep[see also][]{tuomi2012e}, it appears that accounting for correlations on the time-scale of $\sim$ 10 days might disable the detectability of some signals in RV data. When the significance of a signal depends on whether one uses a correlated noise term in the statistical model, it might not be possible to tell if the signal is a genuine one unless the chosen noise model that enables the detection produces a much better solution than the alternative noise models. Regarding GJ 163, the detection of a possible companion at 125 days is not possible if an MA component is included in the noise model (the corresponding signal is completely absorbed by the correlation term and the Markov chains do not converge to a fourth periodicity). On the other hand, a simpler white noise model enables a convergence to a clear four-Keplerian solution and provides a substantially higher model probability. For this data set, it looks like the MA model is over-parameterised and that the inclusion of a fourth periodic signal would be our preferred model choice. Therefore, the best noise model depends on the number of signals and the preferred number of signals depends on the selected noise model. When this happens, the only way to tell what is the most probable situation in reality is to examine all realistically possible combinations (white noise only, red and white noise, $k=0, 1, ...$) and derive the most informed result from there.

The liquid-water habitable zone (HZ) of GJ 163 is located between roughly 0.123 and 0.275 AU \citep[assuming 0\% clouds in the equations of][]{selsis2007}. This suggests that GJ 163 c, with a semi-major axis of 0.126 AU and a likely circular orbit, is located inside the stellar HZ throughout most of its orbital cycle. The minimum mass of GJ 163 c is $m_{p} \sin i =$ 8.7 [5.3, 12.3] M$_{\oplus}$ and its expected true value ($\sim$ 13.0 M$_{\oplus}$) lies therefore roughly in the Neptune-mass regime. If the planet were found to transit in front of the star, the minimum mass would be its true mass, and its size could be obtained and nature elucidated (scaled-down version of Neptune rather or a massive rocky planet with a solid surface). It is therefore a primary target for transit follow-up observations. Given the uncertainties and detailed modelling necessary to account for all the unknowns, assessing the habitability of GJ 163 c in detail is beyond the scope of this work.

The detection of three (possibly four) planet candidates adds GJ 163 to the emerging population of diverse planetary systems around low-mass stars. The architecture of the GJ 163 system resembles a scaled-down version of the Solar System in the sense that the system consists of two super-Earth mass objects with minimum masses of 9.9 and 8.7 M$_{\oplus}$ and short orbital periods of 8.6 and 25.6 days, respectively; a possible cool Neptune with a minimum mass of 14.0 M$_{\oplus}$ and an orbital period of 125 days; and a more massive sub-Saturnian outer planet with an orbital period of 572 days.

\begin{acknowledgements}
M. Tuomi acknowledges D. Pinfield and RoPACS (Rocky Planets Around Cool Stars), a Marie Curie Initial Training Network funded by the European Commission's Seventh Framework Programme. G. Anglada-Escud\'e is supported by the German Federal Ministry of Education and Research under 05A11MG3. Based on data obtained from the ESO Science Archive Facility under request number ANGLADA36104. This research has made use of the SIMBAD database, operated at CDS, Strasbourg, France. The authors acknowledge the significant efforts of the HARPS-ESO team in improving the instrument and its data reduction pipelines that made this work possible. We also acknowledge the efforts of all the individuals that have been involved in observing the target star with HARPS.
\end{acknowledgements}


\end{document}